\documentclass[smallextended]{svjour3}
\smartqed

  \usepackage{color}
  \usepackage{newlfont}
  \usepackage{graphicx}  
  \usepackage{amssymb}
  \usepackage{amsmath,mathrsfs}
  \usepackage{verbatim}
  \usepackage[latin1]{inputenc}
  \usepackage{bbm}
  \usepackage{multirow}
  \usepackage[thinlines]{easytable}
  \usepackage{braket}
  \usepackage[varg]{txfonts}
\usepackage{xcolor}

\newcommand{\be}{\begin{equation}}
\newcommand{\ee}{\end{equation}}
\newcommand{\beq}{\begin{eqnarray}}
\newcommand{\eeq}{\end{eqnarray}}
\newcommand{\Tr}{\text{Tr}}

\begin{document}

\title{Information-based approach towards a unified resource theory}


\author{A. C. S. Costa         \and
        R. M. Angelo
}


\institute{Corresponding Author: A. C. S. Costa \\
              ana.sprotte@gmail.com \\ \\
           R. M. Angelo \\
              renato@fisica.ufpr.br \\ \\
              Department of Physics, Federal University of Paran\'a, P.O. Box 19044, Curitiba, Paran\'a 81531-980, Brazil 
}

\date{Received: date / Accepted: date}

\maketitle

\begin{abstract}
Resource theories play an important role in quantum information theory, as they identify resourceful states and channels that are potentially useful for the accomplishment of tasks that would be otherwise unreachable. The elementary structure of such theories, which is based on the definition of free states and free operations, successfully accommodates different nonclassical aspects, such as quantum coherence and entanglement, but it is still not clear whether and how far such formal framework can be extended. In this work, by taking information as the most primitive quantum resource and defining resource-destroying operations, we develop a unifying approach that proves able to encompass several nonclassical aspects, including the newly developed concepts of quantum irreality and realism-based nonlocality.
\end{abstract}

\section{Introduction}
\label{intro}
Resource theory is a formal framework developed to investigate and analyse the properties of physical systems that can be viewed as resources for certain tasks under restrictive operations~\cite{chitambar2019}. The most common example of a quantum resource is entanglement~\cite{horodecki2009}, which is the basic element of many quantum information processing tasks and formally emerges as a distinctive operational advantage when one is restricted to local operations and classical communication. The basic structure of a resource theory relies on the definition of the following elements: (i) free states, (ii)~resource states, and (iii) free operations arising from natural restrictions on physical systems. In addition, within this formal structure, it is naturally required that no resource state can be created from any free state via free operations.  

The resource theory framework has been applied to various concepts of quantum information science, such as entanglement~\cite{horodecki2009}, purity~\cite{horodecki2003}, coherence~\cite{baumgratz2014,winter2016}, informational equilibrium in quantum thermodynamics~\cite{gour2015,goold2016}, and quantum discord~\cite{liu2017}. The search for a unifying framework of resource theories has attracted much interest in the last few years~\cite{brandao2015,horodecki2013,coecke2016,liu2017,takagi2019,regula2019}. In particular, in Ref.~\cite{liu2017}, the authors developed a general scheme based on resource-destroying maps for the analysis of resource theories. These maps have the property of maintaining free states unchanged and suppressing all the resource from resourceful states. 

Recently, new quantifiers of nonclassicality were introduced with basis on an operational scheme to assess elements of reality~\cite{bill2015}. In a protocol involving preparation, unrevealed measurements, and quantum state tomography, a quantifier for the degree of irreality of an observable for a given state preparation was introduced. Then, as a measure of alterations in the irreality of an observable $A$ induced by unrevealed measurements of observable $B$ conducted in a remote site, the notion of realism-based nonlocality was introduced and their main characteristics studied~\cite{gomes2018,gomes2019}. Whether these concepts admit a formalization as quantum resources remains an important open question.

This work aims at constructing a unifying framework that can apply to several well-established nonclassical aspects and is sufficiently elastic to accommodate future concepts. To this end, we first review the formal aspects underlying the resource theory of information and introduce the destroying monitoring---a map that interpolates between the regime in which no resource is generated and the one where absolutely all information is suppressed. Then, we show by explicit construction that quantum coherence, entanglement, one-way and symmetric quantum discord, quantum irreality, and realism-based nonlocality can all be framed in terms of a picture in which an external agent extracts information about some observable of the system. Interestingly, our approach implies that the suppression of all these quantum resources can be viewed as a process that establishes realism.

\section{Preliminaries}
\subsection{Resource theory \label{RT}}
A given nonclassical feature $\mathcal{R}$ is termed a {\it resource} whenever the following formal structure can be maintained. Let $\mathcal{R}(\rho)$ be a reasonable nonnegative quantifier of the amount of such resource encoded in the state $\rho\in\mathcal{B(H)}$, where $\mathcal{B(H)}$ is the set of bounded operators acting on the Hilbert space $\mathcal{H}$. Let $\mathcal{F}\subset \mathcal{B(H)}$ be the set of free states $\tilde{\rho}$. This means that for all $\tilde{\rho}\in\mathcal{F}$ one has $\mathcal{R}(\tilde{\rho})=0$. States $\rho_{\max}$ such that $\mathcal{R}(\rho_{\max})=\mathcal{R}_{\max}$ are called maximum-resource states. Hence, $\rho$ is a resource state if $0<\mathcal{R}(\rho)\leqslant \mathcal{R}_{\max}$. Now, let the completely positive trace-preserving (CPTP) map $\Theta$ ($\Phi$) be a quantum operation that suppresses none of (all) the resource $\mathcal{R}$ available in $\rho$. That is, $\mathcal{R}(\Theta(\rho))=\mathcal{R}(\rho)$ and $\mathcal{R}(\Phi(\rho))=0$. Accordingly, $\Theta$ is said a nongenerating operation,
whereas $\Phi$ is a maximally destroying one. For the sake of simplicity and generality, here we set $\Theta=\mathbbm{1}$, which thus plays the role of a universal nongenerating operation. We now introduced an operation $\Lambda_\epsilon$, which will throughout be referred to as {\it monitoring}. Mathematically, it is represented by the CPTP map
\be
\Lambda_\epsilon(\rho)=(1-\epsilon)\,\rho+\epsilon\,\Phi(\rho),
\label{monitoring}
\ee
where the parameter $\epsilon\in[0,1]$ is introduced to characterize the capability of the monitoring in destroying resource. In effect, one has $\mathcal{R}(\Lambda_0(\rho))=\mathcal{R}(\rho)$ and $\mathcal{R}(\Lambda_1(\rho))=0$. If $\mathcal{R}$ is a convex measure, that is, $\mathcal{R}\left(\sum_ip_i\rho_i\right)\leqslant \sum_ip_i\mathcal{R}(\rho_i)$, then we find 
\be
\mathcal{R}(\Lambda_\epsilon(\rho))\leqslant \mathcal{R}(\rho), 
\label{monotonicity}
\ee 
equality holding only for $\epsilon=0$. Therefore, we see that, being able to implement a smooth interpolation between the extreme regimes of no resource destruction whatsoever ($\epsilon=0$) and total destruction ($\epsilon=1$), the monitoring $\Lambda_\epsilon$ turns out to formally be a generic description of a free operation. In particular, via convexity, it is guaranteed that $\mathcal{R}(\Lambda_\epsilon(\tilde{\rho}))=0$. 

Altogether the above statements set a basic formal structure of a resource theory for a general resource $\mathcal{R}$. Upon specialization of the resource of interest, one may proceed to define the monitoring $\Lambda_\epsilon$, the convex quantifier, and the particular quantum information tasks for which resourceful states yield operational advantage with respect to free states.

\subsection{Information as a resource}
\label{inf-resource}
Here, we review the well-established structure that renders information the status of a resource. This is opportune because, as shown later, information can be adopted as a primitive resource from which many others derive. Being dual to the ignorance measured via the von Neumann entropy $S(\rho)$, information is quantified as~\cite{horodecki2003}
\be 
I(\rho)\coloneqq \ln{d}-S(\rho),
\ee 
where $\rho\in\mathcal{B(H)}$ and $d=\dim\mathcal{H}$. Besides being connected with work extraction~\cite{horodecki2005} and playing a distinctive role in the resource theory of purity, it proves to be the unique measure of information in the context of reversible transformations from pure to mixed states~\cite{horodecki2003}. Furthermore, it is related to the maximal quantum coherence~\cite{streltsov2018}. 

Information directly fits in the above delineated formal structure. The demonstration goes as follows. It is clear that $\tilde{\rho}=\mathbbm{1}/d$ and $\rho_{\max}=\ket{\psi}\bra{\psi}$ are the free state and the maximum-resource state, respectively. The maximally destroying operation is constructed as follows. Let us henceforth consider the composite space $\mathcal{H=H_A\otimes H_B}$ with $A=\sum_aaA_a\in\mathcal{B(H_A)}$ being a discrete-spectrum observable with projectors $A_a=\ket{a}\bra{a}$. Also, we consider the so-called unrevealed measurement map,
\be
\Phi_A(\rho)\coloneqq \sum_a \left(A_a\otimes\mathbbm{1}_\mathcal{B}\right)\,\rho\,\left(A_a\otimes\mathbbm{1}_\mathcal{B}\right),
\label{PhiA}
\ee 
which is associated with nonselective projective measurements of the observable $A$~\cite{bill2015}. Since, in general, the von Neumann entropy is increasing under CPTP maps, it follows that $I(\Phi_A(\rho))\leqslant I(\rho)$, so that $\Phi_A$ is a free operation, although not the maximally destroying one. Similar considerations apply for $\Phi_B$, which refers to unrevealed measurements of $B\in\mathcal{B(H_B)}$. If $\{A,A'\}\in\mathcal{B(H_A)}$ are maximally incompatible observables, then $\Tr(A_aA'_{a'})=1/d_\mathcal{A}$ and it is not difficult to show that $\Phi_{AA'}(\rho)\equiv\Phi_A\Phi_{A'}(\rho)=\frac{\mathbbm{1}_\mathcal{A}}{d_\mathcal{A}}\otimes\rho_\mathcal{B}$, where $\rho_\mathcal{B}=\Tr_\mathcal{A}(\rho)$. Note that $\Phi_{AA'}$ is not the maximally destroying operation yet, since $I(\Phi_{AA'}(\rho))=I(\rho_\mathcal{B})$. The searched operation is $\Phi_{AA'}\Phi_{BB'}$, for maximally incompatible observables $\{B,B'\}\in\mathcal{B(H_B)}$, since in this case we find $\Phi_{AA'}\Phi_{BB'}(\rho)=\mathbbm{1}/d$ (with $d=d_\mathcal{A}d_\mathcal{B}$), which has no residual information whatsoever. Therefore, as far as information is concerned, the resource-destroying monitoring specializes as
\be
\Lambda_\epsilon(\rho)=(1-\epsilon)\,\rho+\epsilon\,\Phi_\text{inc}(\rho),
\label{Lambda_I}
\ee
where $\Phi_\text{inc}\equiv\Phi_{AA'}\Phi_{BB'}$ is the pairwise map involving incompatible observables. Via monotonicity of the von Neumann entropy, $S(\Lambda_\epsilon(\rho))\geqslant S(\rho)$, one proves that $I(\Lambda_\epsilon(\rho))\leqslant I(\rho)$, which shows that the above map correctly describes a general free operation. Moreover, we have $I(\Lambda_0(\rho))=I(\rho)$ and $I(\Lambda_1(\rho))=0$.

\subsection{Conservation of the total resource}
\begin{figure}[htp]
\begin{center}
\includegraphics[scale=0.4]{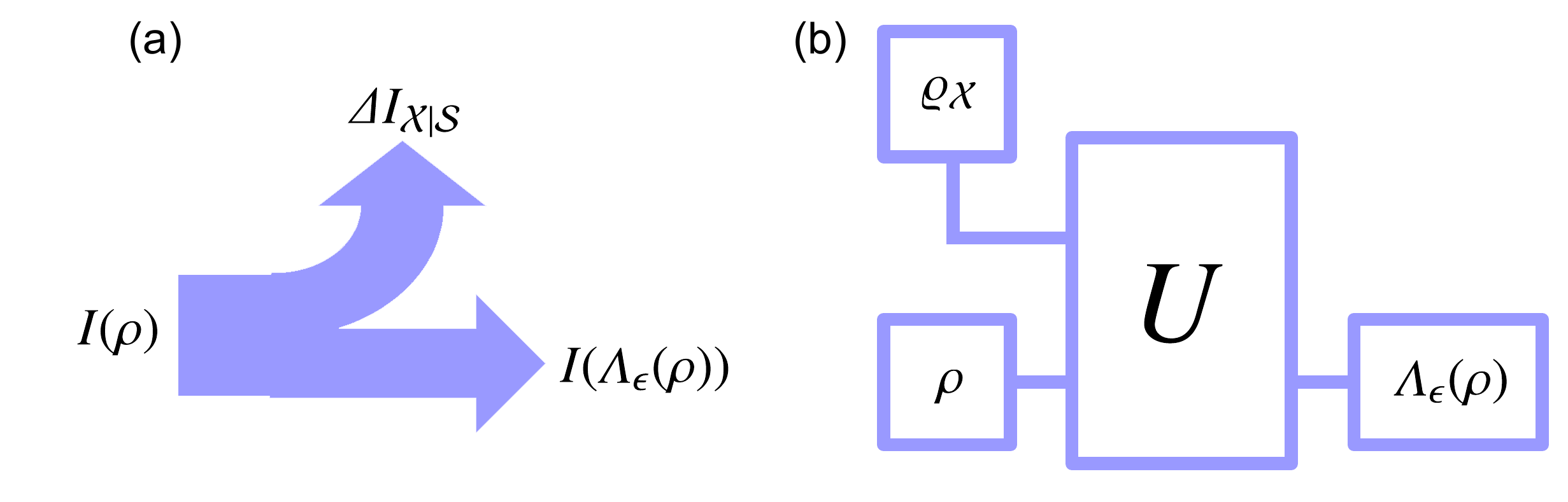}
\caption{Schematic diagrams for the information flow. Depictions are given in (a) the informational space and in (b) the state space. The transition $\rho\mapsto \Lambda_\epsilon(\rho)$ occurs after the system gets correlated, via a unitary operation $U$, with a (discarded) system $\mathcal{X}$. The amount $\Delta I_\mathcal{X|S}=I(\rho)-I(\Lambda_1(\rho))$ of resource suppressed from the system is delivered to the external agent $\mathcal{X}$.}
\label{fig1}
\end{center}
\end{figure}

The informational resource suppressed due to the action of a resource-destroying monitoring is not banished from the universe; it is just encoded in another sector of the Hilbert space. In fact, the resource is extracted by an external agent (an auxiliary system henceforth denoted $\mathcal{X}$) that gets information about some observable of the system. By use of the Stinespring dilation theorem~\cite{nielsen2000}, one shows that any quantum operation $\Lambda_\epsilon$ on $\rho\in\mathcal{B(H)}$, with $\mathcal{H=H_A\otimes H_B}$, can be viewed as a reduced unitary evolution of the system $\mathcal{S\equiv AB}$ coupled to $\mathcal{X}$, that is,
\be
\Lambda_\epsilon(\rho)=\Tr_\mathcal{X} \Big[U\,\varrho(0)\,U^\dagger\Big]=\varrho_\mathcal{S}(t),
\label{Lambda}
\ee
where $\varrho(0) = \rho \otimes\ket{x_0}\bra{x_0}$ and $U$ is a unitary operation acting on $\mathcal{H\otimes H_X}$. Rewriting the mutual information, $I_\mathcal{X:S}(\varrho(t))=S(\varrho_\mathcal{S}(t))+S(\varrho_\mathcal{X}(t))-S(\varrho(t))$ between the parts $\mathcal{S}$ and $\mathcal{X}$, with $\varrho(t)=U\,\varrho(0)\,U^{\dag}$ and $\varrho_\mathcal{X}(t)=\Tr_\mathcal{S}[\varrho(t)]$, in terms of information, produces $I(\varrho(t))=I(\varrho_\mathcal{S}(t))+I(\varrho_\mathcal{X}(t))+I_\mathcal{X:S}(\varrho(t))$. With Eq.~\eqref{Lambda}, this expression reduces to
\be
I(\varrho(t))=I(\Lambda_\epsilon(\rho))+I(\varrho_\mathcal{X}(t))+I_\mathcal{X:S}(\varrho(t)).
\label{Itot}
\ee 
Given that the state of $\mathcal{X}$ is initially pure and there is no correlation between $\mathcal{X}$ and $\mathcal{S}$, at $t=0$, we have
\be
I(\varrho(0)) = I(\rho) + \ln{d_\mathcal{X}},
\ee
Now, consider the information that $\mathcal{X}$ encodes about $\mathcal{S}$, 
\be 
I_\mathcal{X|S}(\varrho(t))=\ln{d_\mathcal{X}}-S_\mathcal{X|S}(\varrho(t)),
\ee 
where $S_\mathcal{X|S}(\varrho(t))=S(\varrho(t))-S(\varrho_\mathcal{S}(t))$ is the conditional entropy of $\mathcal{X}$. Note that $I_\mathcal{X|S}(\varrho(0))=\ln{d_\mathcal{X}}$. Using the unitary invariance of the von Neumann entropy, which implies that $I(\varrho(t))=I(\varrho(0))$, we combine the above relations to arrive at
\be
I(\rho)=I(\Lambda_\epsilon(\rho)) + \Delta I_\mathcal{X|S},
\label{inf-flow}
\ee
where the term $\Delta I_\mathcal{X|S}=I_\mathcal{X|S}(\varrho(t))-I_\mathcal{X|S}(\varrho(0))$ can be read as the increment of information storage about $\mathcal{S}$ in $\mathcal{X}$. In fact, this factor can be rewritten in the form $\Delta I_\mathcal{X|S}=\Delta I_\mathcal{X}(t)+\Delta I_{\mathcal{X:S}}(t)$, with $\Delta I_\mathcal{X}(t)=I(\varrho_\mathcal{X}(t))-\ln{d_\mathcal{X}}$ and $\Delta I_\mathcal{X:S}(t)=I_\mathcal{X:S}(t)$. This shows that the decreasing of informational resource induced in the process $\rho\mapsto \Lambda_\epsilon(\rho)$ is accompanied by a flow of conditional information to another part of the (dilated) system. Figure \ref{fig1} depicts such resource conservation. Also noteworthy in this discussion, for instance from the decomposition \eqref{Itot}, is the idea that mutual information itself is a mode in which information is codified, being therefore informational resource as well. The difference, though, is that it is spread over two parts.

In what follows, we show that several well-known resources can be cast as a flow $\Delta I_\mathcal{X|S}=I(\rho)-I(\Lambda_1(\rho))$ induced by some maximally destroying monitoring $\Lambda_1$. The description of each particular resource is shown to emerge from the specification of $\Lambda_1$, which is intrinsically associated, via particularly chosen unitary transformations $U$ [see Eq.~\eqref{Lambda}], with the observable about which the external agent $\mathcal{X}$ extracts information. 

As a side remark, we note that, in more general contexts, the strength $\epsilon$ of the monitoring $\Lambda_\epsilon$, when equipped with~\eqref{PhiA}, can be linked with weak measurements~(see Ref.~\cite{dieguez2018} for an explicit construction). In addition, the monitoring can be framed in the operator-sum representation~\cite{nielsen2000} as $\Lambda_\epsilon (\rho) = \sum_a K_a \rho K_a^\dagger$, with $K_0^\dagger K_0+\sum_a K_a^\dagger K_a=\mathbbm{1}$, $K_0 = \sqrt{1-\epsilon}\,\mathbbm{1}$, and $K_a=\sqrt{\epsilon}\,(A_a\otimes\mathbbm{1}_{\mathcal{B}})$. Also note that the maps $\Lambda_\epsilon$ employed in the previous sections are particular forms of the one defined in Eq.~\eqref{Lambda} and used throughout this section.

\section{Unifying perspective}
Now we are ready to introduce our unifying approach to quantum resources. Basically, here we show that quantum resources can be interpreted as the amount of information that flows off due to monitorings conducted by an external agent, where the maximally destroying map $\Phi$ has to be specified in each case.

\subsection{Quantum coherence}
We start with quantum coherence~\cite{streltsov2017a}. For a given fixed basis $\{\ket{a}\}$, associated with an observable $A=\sum_aaA_a\in\mathcal{B(H_A)}$, with projectors $A_a=\ket{a}\bra{a}$, we have
\be
\tilde{\rho}=\sum_a^{d_\mathcal{A}} p_a \ket{a}\bra{a} \quad\textrm{and}\quad \rho_{\max} = \frac{1}{d_\mathcal{A}}\sum_{a,a'}^{d_\mathcal{A}}\ket{a}\bra{a'}
\label{free-max}
\ee
for the free states (the so-called incoherent states) and the maximum-resource states, respectively. The distillable coherence~\cite{baumgratz2014,winter2016}, given by the relative entropy of coherence, $C_A(\rho)$ of $\rho\in\mathcal{B(H_A)}$, has the following closed expression
\be
C_A(\rho)=I(\rho)-I(\Phi_A(\rho)).
\label{C_A}
\ee
Note that if $\rho$ is an incoherent state (with respect to the fixed basis), then $\rho=\Phi_A(\rho)\coloneqq\sum_aA_a\rho A_a$ and $C_A(\rho)=0$, as is desirable. Inspired by the framework discussed in the previous section, we see via definition \eqref{C_A} that the coherence $C_A$ of a state $\rho$ is nothing else than the amount of informational resource that would flow off upon the monitoring $\Lambda_1=\Phi_A$. In this sense, information emerges as a primitive resource underlying coherence. 

Of course, quantum coherence has shown to be a resource according to the formal structure defined by the states \eqref{free-max}, the quantifier \eqref{C_A}, and the known free operations (incoherent operations)~\cite{baumgratz2014}, which turn out to be equivalent to a coherence destroying monitoring $\Lambda_\epsilon^{C_A}(\rho)=(1-\epsilon)\,\rho+\epsilon\,\Phi_A(\rho)$. In effect, from the concavity of von Neumann's entropy and the fact that $\Phi_A(\Lambda_\epsilon^{C_A}(\rho))=\Phi_{A}\Phi_{A}(\rho)=\Phi_{A}(\rho)$, one can straightforwardly check that $C_A(\Lambda_\epsilon^{C_A}(\rho))\leqslant C_A(\rho)$, with equality holding only for $\epsilon=0$ or $\rho=\Phi_A(\rho)$.

Summarizing, for coherence we have an explicit structure equipped with free states, maximum-resource states, a nonnegative quantifier, and a mimic in terms of information flow. So far, we have been dealing with single-partite states. In what follows, our discussion extends to bipartite cases, where quantum correlations emerge as the fundamental resources.

\subsection{Entanglement}
Now, let us move our focus to entanglement, which is widely recognized as an important resource~\cite{horodecki2009} for tasks such as quantum cryptography~\cite{gisin2002}, randomness generation~\cite{pironio2010}, and quantum metrology~\cite{toth2014}. Within the resource theory, entanglement stands out whenever the framework is restricted to local operations and classical communication. 

Let $E(\psi)$ denote the entanglement of a pure bipartite state $\ket{\psi}$. Consider the Schmidt operator $\Gamma_\mathcal{A}=\sum_i\gamma_i\ket{\gamma_i}\bra{\gamma_i}$ acting on $\mathcal{H_{A}}$, such that $\ket{\psi}=\sum_i\sqrt{\lambda_i}\ket{\gamma_i}\ket{\gamma_i}$. As is well known, the natural quantifier of entanglement, in this case, is the von Neumann entropy of the reduced state, that is, $E(\psi)=S(\Tr_\mathcal{B}\ket{\psi}\bra{\psi})=H(\{\lambda_i\})$, where $H(\{\lambda_i\})$ is the Shannon entropy of the Schmidt probability distribution $\lambda_i$. It can be straightforwardly checked that this measure can be expressed in the form
\be
E(\psi)=\mathcal{I}(\psi)-\mathcal{I}\big(\Phi_{\Gamma_\mathcal{A}}(\psi)\big),
\label{E(psi)}
\ee 
where $\mathcal{I}(\psi)\equiv\mathcal{I}(\ket{\psi}\bra{\psi})$, $\mathcal{I}\in\{I,I_\mathcal{A:B}\}$, and $\Phi_{\Gamma_\mathcal{A}}(\psi) \equiv \Phi_{\Gamma_\mathcal{A}}(\ket{\psi}\bra{\psi})$. This relation makes it explicit that pure-state entanglement can be primarily viewed, in accordance with Eq.~\eqref{inf-flow}, as an informational resource (actually, either information or mutual information) that is subtracted via destroying monitoring of a Schmidt observable. By use of the entanglement of formation~\cite{bennett1996,wootters1998}, the extension of this idea to a mixed state $\rho=\sum_ip_i\ket{\psi_i}\bra{\psi_i}$ is straightforward: 
\be
E(\rho)=\min_\mathscr{E}\sum_ip_i\Big[\mathcal{I}(\psi_i)-\mathcal{I}(\Phi_{\Gamma_\mathcal{A}^i}(\psi_i))\Big],
\label{E(rho)}
\ee 
where the minimization runs over all ensembles $\mathscr{E}=\{p_i,\ket{\psi_i}\}$ through which $\rho$ can be decomposed. The above formula allows us to interpret entanglement as the average amount of informational resource that is removed from an optimal ensemble (which allows the minimization) via unrevealed measurements of the local Schmidt operator $\Gamma_\mathcal{A}^i$.

For pure-state entanglement, the resource-destroying map can be written as a monitoring of the form \eqref{monitoring}. The resulting map reads $\Lambda_\epsilon^E (\psi) = (1-\epsilon)\,\ket{\psi}\bra{\psi} +\epsilon\,\Phi_{\Gamma_\mathcal{A}}(\psi)$. Using entropy concavity and the fact that $\Phi_{\Gamma_\mathcal{S}}(\Lambda_\epsilon^E(\psi))=\Phi_{\Gamma_\mathcal{A}}(\psi)$ one checks that $E(\Lambda_\epsilon^E(\psi))\leqslant E(\psi)$. In particular, one has $E(\Lambda_1^E(\psi))=0$. A similar formulation for the mixed state case is, however, much harder to conceive because in this case an entire set $\{\Gamma_\mathcal{A}^i\}$ of Schmidt operators has to be specified, while the structural form employed here for the monitoring admits only one operator. Free states and maximum-resource states are well known, so that here as well we have a complete formal structure supplemented with an information-based interpretation.

\subsection{Quantum discord}
Quantum discord~\cite{zurek2001,vedral2001} has shown to be a useful resource for tasks in quantum computation~\cite{datta2008}, quantum information processing~\cite{dakic2012}, quantum communication~\cite{madhok2013}, and quantum cryptography~\cite{pirandola2014}. Although being a resource for different tasks, its formalization within the scope of a resource theory remains open due to its nonconvexity and the lack of a clear understanding of its free operations structure~\cite{liu2017}. Here, we show that quantum discord can be naturally accommodated in our unifying framework.

The one-way quantum discord $\mathcal{D_A}$, as rephrased by Rulli and Sarandy~\cite{rulli2011}, is written as $\mathcal{D_A}(\rho)=\min_AD_A(\rho)$ with
\be
D_A(\rho)=I_\mathcal{A:B}(\rho)-I_\mathcal{A:B}(\Phi_A(\rho)).
\label{D_A}
\ee
Similar expressions apply for $\mathcal{D_B}$. The quantity $D_A$ is the so-called basis-dependent quantum discord, here referred to as the $A$-discord, for simplicity. This formulation allows us to interpret quantum discord as the minimum amount of mutual information (a resource) that flows to an external agent that monitors the observable $A\in\mathcal{B(H_A)}$. $A$-discord can be explicitly written in terms of information as
\be
D_A(\rho)=\big[I(\rho)-I(\Phi_A(\rho))\big]-\big[I(\rho_{\mathcal{A}})-I(\Phi_A(\rho_{\mathcal{A}}\otimes\mathbbm{1}_{\mathcal{B}}))\big].
\ee
Here, the interpretation is a bit subtler. $A$-discord can be viewed as the informational surplus that occurs when a maximally destroying monitoring $\Lambda_1=\Phi_A$ is realized in two independent processes, one involving the joint state $\rho$ and another involving the reduced state $\rho_\mathcal{A}$. Alternatively, by noting that $\Phi_{BB'}(\rho)=\rho_\mathcal{A}\otimes\frac{\mathbbm{1}}{d_\mathcal{B}}$ and, as a consequence, that $I(\Phi_{BB'}(\rho))=I(\rho_\mathcal{A})$, we arrive at
\be
D_A(\rho)=\big[I(\rho)-I(\Phi_A(\rho))\big]-\big[I(\varrho)-I(\Phi_A(\varrho))\big]
\ee
where $\varrho\equiv\Phi_{BB'}(\rho)$. Here, we have $A$-discord as the extra information in the comparison of a destroying monitoring involving $\rho$ and another involving a previously monitored state $\Phi_{BB'}(\rho)$, with maximally incompatible observables $\{B,B'\}$. A further interpretation can be built 
by recalling that other maps exist which can change the informational resource of a state. A distinctive one is the partial trace (PT), which maps a density operator on $\mathcal{H_A\otimes H_B}$ onto a density operator on $\mathcal{H_A}$. It can be implemented by the use of the Kraus operators $\text{\small $M_k$}=\mathbbm{1}_\mathcal{A}\otimes\bra{b_k}$ satisfying $\sum_k\text{\small $M_k^\dag M_k$}=\mathbbm{1}_\mathcal{AB}$ and the operator-sum representation, which gives $\Lambda_\text{PT}(\rho)=\sum_k \text{\small $M_k$}\rho \text{\small $M_k^\dag$}=\rho_\mathcal{A}$. With this observation, we can rewrite the conditional information $I_\mathcal{B|A}(\rho):=\ln{d_\mathcal{B}}-S_\mathcal{B|A}(\rho)=I(\rho)-I(\rho_\mathcal{A})$, where $S_\mathcal{B|A}(\rho)=S(\rho)-S(\rho_\mathcal{A})$, in the form  $I_\mathcal{B|A}(\rho)=I(\rho)-I(\Lambda_\text{PT}(\rho))$. It follows, therefore, that the conditional information is itself an informational resource since it is the amount of information that flows off via a generalized quantum operation $\Lambda_\text{PT}$. This allows us to write $A$-discord as
\be 
D_A(\rho)=I_\mathcal{B|A}(\rho)-I_\mathcal{B|A}(\Phi_A(\rho)),
\ee  
which indicates a monitoring-based flow of conditional information. All this fits in our unifying perspective.

In Ref.~\cite{rulli2011}, the authors extended their formulation to the symmetric quantum discord $\mathcal{D}(\rho)=\min_{A,B}D_{A,B}(\rho)$, where
\be
D_{A,B}(\rho)=I_\mathcal{A:B}(\rho)-I_\mathcal{A:B}(\Phi_A\Phi_B(\rho)).
\ee
Henceforth termed $\{A,B\}$-discord, $D_{A,B}$ can also be rephrased in terms of information, the resulting expression being 
\beq
D_{A,B}(\rho)&=&\big[I(\rho) - I(\Phi_A\Phi_B(\rho))\big]-\big[I(\rho_{\mathcal{A}}) - I(\Phi_A(\rho_{\mathcal{A}}\otimes\mathbbm{1}_{\mathcal{B}}))\big] \nonumber \\
&-&\big[I(\rho_{\mathcal{B}}) - I(\Phi_B(\mathbbm{1}_{\mathcal{A}}\otimes\rho_{\mathcal{B}}))\big].
\eeq
Thus, $\{A,B\}$-discord can be viewed as the informational surplus deriving from the balance among three independent processes involving the destroying monitorings $\{\Phi_A,\Phi_B,\Phi_A\Phi_B\}$ and the states $\{\rho_\mathcal{A},\rho_\mathcal{B},\rho\}$. Alternatively, one may write 
\be 
D_{A,B}(\rho)=D_A(\rho)+D_B(\Phi_A(\rho)),
\ee 
which rephrases $\{A,B\}$-discord in terms of the corresponding one-way resources.

The maximum-resource state in the discord context is expected to be the maximally entangled state since quantum discord is equivalent to entanglement for pure states (instance in which we have a high level of quantum correlations). Stronger evidence to this claim is provided in Ref.~\cite{xi2012}, where the authors show that pure states are the only ones that saturate the inequality $\mathcal{D_A}(\rho) \leqslant S(\rho_{\mathcal{A}})$ \cite{xi2011}. Free states, by their turn, are given by $\tilde{\rho}=\Phi_A(\varrho)$, for any state $\varrho$. 

For the discussion of the discord monitoring, let $\bar{A}$ be the observable that implements the $A$-discord minimization so that $\mathcal{D_A}(\rho)=D_{\bar{A}}(\rho)$. Although candidates for discord nongenerating operations have already been reported, such as the unitary-isotropic channels~\cite{liu2017}, here we keep considering $\Theta=\mathbbm{1}$, for the sake of unification. We then propose the map $\Lambda_\epsilon^\mathcal{D_A}(\rho)=(1-\epsilon)\,\rho+\epsilon\,\Phi_{\bar{A}}(\rho)$. Because $\Phi_{\bar{A}}\Lambda_\epsilon^\mathcal{D_A}=\Phi_{\bar{A}}$ and mutual information is monotonic under CPTP maps, one proves that $\mathcal{D_A}(\Lambda_\epsilon^\mathcal{D_A}(\rho))\leqslant \mathcal{D_A}(\rho)$, with equality holding for $\epsilon=0$ and $\rho=\Phi_{\bar{A}}(\rho)$. For the symmetric quantum discord, one can propose, by means of a similar rationale, the monitoring $\Lambda_\epsilon^\mathcal{D}(\rho)=(1-\epsilon)\,\rho+\epsilon\,\Phi_{\bar{A}}\Phi_{\bar{B}}(\rho)$, which leads to $\mathcal{D}(\Lambda_\epsilon^\mathcal{D}(\rho))\leqslant \mathcal{D}(\rho)$ and $\mathcal{D}(\Lambda_1^\mathcal{D}(\rho))=0$. Therefore, once again, we succeeded to apply our unifying framework to a well-established nonclassical aspect. With little effort, the same can be accomplished for the recently introduced ``weak versions'' of quantum discord~\cite{dieguez2018}.
 
\subsection{Quantum irreality}
 
In the past few years, the condition $\rho=\Phi_A(\rho)$ has been adopted in some frameworks~\cite{bill2015} as a criterion of reality for the observable $A$, meaning that for such a preparation $A$ is an element of reality. This criterion alone allows us to ascribe a different interpretation for the monitoring (see Eq.~\eqref{monitoring} and Fig.~\ref{fig1}): it can be viewed as an action intended to increase the degree of reality of $A$. Then, in accordance with Eq.~\eqref{inf-flow}, $\Delta I_\mathcal{X|S}=I(\rho)-I(\Lambda_1(\rho))$ gives the amount of information extracted by $\mathcal{X}$ to make $A$ become an element of reality.

The degree of irreality $\mathfrak{I}_A(\rho)$ of an observable $A$ for an arbitrary preparation $\rho$ is introduced as $\mathfrak{I}_A(\rho)=S(\rho||\Phi_A(\rho))$, which is shown to reduce to $\mathfrak{I}_A(\rho)=S(\Phi_A(\rho))-S(\rho)$. Clearly, this quantity estimates the ``entropic distance'' between $\rho$ and the $A$-reality state $\Phi_A(\rho)$. Irreality is known to be nonnegative and vanishing if and only if $\rho=\Phi_A(\rho)$. This notion of ``quantum irrealism'' has recently been explored for different perspectives, such as for the statement of an information-reality complementarity relation~\cite{dieguez2018a}, the discussion of continuous-variable realism~\cite{freire2019}, the identification of quantum irreality in quantum walks~\cite{orthey2019}, and for the formulation of the realism-based nonlocality~\cite{bill2015,gomes2018,gomes2019}, which will be discussed in the next section. In terms of information, the irreality measure reads
\be
\mathfrak{I}_A(\rho)=I(\rho)-I(\Phi_A(\rho)).
\label{frakI}
\ee
It is then manifest that irreality admits the identification with the amount of informational resource that flows off upon unrevealed measurements of $A$. Also, as shown in Ref.~\cite{bill2015}, irreality can be decomposed in the form
\be
\mathfrak{I}_A(\rho)=\mathfrak{I}_A(\rho_\mathcal{A})+D_A(\rho),
\label{LI+D}
\ee
where $\mathfrak{I}_A(\rho_\mathcal{A})$ is nothing but the relative entropy of coherence and $D_A(\rho)$ is the aforementioned $A$-discord, both parcels having been already identified as resources. Since $\mathfrak{I}_A(\rho)$ is a clear combination of, say, a local resource (coherence) with a global one (discord), it can be naturally interpreted as a quantum resource as well. In addition, irreality has an evident mathematical equivalence with the tools employed in the context of coherence in distributed scenarios~\cite{chitambar2016,streltsov2017}. In this context, the quantity~\eqref{frakI} is called quantum-incoherent relative entropy and upper bounds the asymptotic rate of assisted coherence distillation under the restriction of local quantum-incoherent operations and classical communication. Furthermore, it can be directly linked with thermal discord and daemonic protocols~\cite{zurek2003}, when a maximization is taken over all possible observables $A$~\cite{costa2013}.

Within the context of an eventual resource theory of quantum irrealism, the trivial candidates for free states can be written as $\tilde{\rho}=\sum_k p_k\Phi_A(\rho^\mathcal{A}_k)\otimes\rho^\mathcal{B}_k$, since it is immediately seen that $\Phi_A(\tilde{\rho})=\tilde{\rho}$ and, therefore, $\mathfrak{I}_A(\tilde{\rho})=0$. To envisage the maximum-resource states for irreality, we search for an upper bound for Eq.~\eqref{frakI}. Using the joint-entropy theorem~\cite{nielsen2000}, we first show that $\mathfrak{I}_A(\rho)=S(\Phi_A(\rho_{\mathcal{A}}\otimes\mathbbm{1}_{\mathcal{B}}))+\sum_a p_a S(\rho_{\mathcal{B}|a}) - S(\rho)$. Then, by means of the inequality $\sum_ap_aS(\rho_{\mathcal{B}|a}) \leqslant S(\rho)$~\cite{xi2011}, we arrive at $\mathfrak{I}_A(\rho) \leqslant S(\Phi_A(\rho_{\mathcal{A}}\otimes\mathbbm{1}_{\mathcal{B}}))$, where the equality holds for pure states. This result implies that irreality is saturated to $\ln d_{\mathcal{A}}$ by maximally entangled states for all observables $A$ and by all pure states whenever the $A$ basis is chosen to be maximally incompatible with the Schmidt basis (with the requirement that $d_{\mathcal{A}} \leqslant d_{\mathcal{B}}$). It then follows that these states turn out to be the maximum-resource states for irreality. 

By virtue of the connection \eqref{frakI} between irreality and information, the resource-destroying map is readily inferred to be $\Lambda_\epsilon^{\mathfrak{I}_A}(\rho)=(1-\epsilon)\,\rho+\epsilon\,\Phi_A(\rho)$. Once again, via monotonicity of the von Neumann entropy, one shows that $\mathfrak{I}_A(\Lambda_\epsilon^{\mathfrak{I}_A} (\rho))\leqslant \mathfrak{I}_A(\rho)$ and $\mathfrak{I}_A(\Lambda_1^{\mathfrak{I}_A} (\rho))=0$. At this point, it becomes clear that processes involving monitorings, such as the generic one depicted in Fig.~\ref{fig1}, typically increase realism.

\subsection{Realism-based nonlocality}
Under the premise that Eq.~\eqref{frakI} is a reasonable quantifier of irreality, the authors of Ref.~\cite{bill2015} proposed a notion of nonlocality that is based on alterations of $A$'s irreality through physical disturbances occurring in a distant site. Specifically, they introduced the contextual realism-based nonlocality
\be
\eta_{A,B}(\rho)=\mathfrak{I}_A(\rho)-\mathfrak{I}_A(\Phi_B(\rho)).
\label{eta}
\ee
This symmetrical (upon the exchange $A\leftrightarrows B$) nonnegative quantity, which vanishes for uncorrelated ($\rho=\rho_\mathcal{A}\otimes\rho_\mathcal{B}$) or reality states [$(\varrho = \Phi_A(\varrho)$ or $\varrho = \Phi_B(\varrho)$], quantifies violations of the hypothesis $\mathfrak{I}_A(\rho)=\mathfrak{I}_A(\Phi_B(\rho))$, according to which the irreality of $A$ in site $\mathcal{A}$ is not affected by unrevealed measurements of $B$ conducted in a remote site $\mathcal{B}$. Maximizing $\eta_{A,B}$ over all contexts $\{A,B\}$ leads to the so-called realism-based nonlocality~\cite{bill2015,gomes2018,gomes2019}, 
\be
\mathcal{N}(\rho)=\max_{\{A,B\}}\eta_{A,B}(\rho),
\label{calN}
\ee
which captures the context-independent realism-based nonlocality, that is, the nonlocal aspects inherent to $\rho$ only. Using the definitions \eqref{frakI} and \eqref{eta} and letting $\{\bar{A},\bar{B}\}$ be the optimal observables such that $\mathcal{N}(\rho)=\eta_{\bar{A},\bar{B}}(\rho)$, one shows that
\be
\mathcal{N}(\rho)=\Big[I(\rho)-I(\Phi_{\bar{A}}(\rho))\Big]-\Big[I(\varrho)-I(\Phi_{\bar{A}}(\varrho))\Big].
\ee
where $\varrho\equiv\Phi_{\bar{B}}(\rho)$. Then, realism-based nonlocality turns out to be the informational surplus that emerges when two maximally destroying monitoring processes related to $\bar{A}$ are compared, one of them involving the state $\rho$ and the other involving the state $\Phi_{\bar{B}}(\rho)$, which presumes a previous monitoring of $\bar{B}$. Interestingly, using the decomposition \eqref{LI+D}, we find
\be
\mathcal{N}(\rho)=D_{\bar{A}}(\rho) - D_{\bar{A}}(\Phi_{\bar{B}}(\rho)),
\ee
which reflects a difference of $\bar{A}$-discords in scenarios defined by $\rho$ and $\Phi_{\bar{B}}(\rho)$. This legitimates us to interpret $\mathcal{N}(\rho)$ as a quantum resource because it is the amount of (another) resource that is suppressed upon a given monitoring.

In the resource theory of realism-based nonlocality, the free states are $\rho_{\mathcal{A}}\otimes\rho_{\mathcal{B}}$, $\Phi_{\bar{A}}(\rho)$ and $\Phi_{\bar{B}}(\rho)$. To find the maximum-resource states for realism-based nonlocality we proceed as follows. First, from $\mathfrak{I}_A(\rho) \leqslant S(\Phi_A(\rho_{\mathcal{A}}\otimes\mathbbm{1}_{\mathcal{B}}))$ and $\mathfrak{I}_A(\Phi_B(\rho))\geqslant 0$, we obtain $\eta_{A,B}(\rho) \leqslant S(\Phi_A(\rho_{\mathcal{A}}\otimes\mathbbm{1}_{\mathcal{B}}))$. Since realism-based nonlocality is symmetric upon the exchange $A\leftrightarrows B$, the inequality $\eta_{A,B}(\rho) \leqslant S(\Phi_B(\mathbbm{1}_{\mathcal{A}}\otimes\rho_{\mathcal{B}}))$ also holds. Then, as $S(\Phi_A(\rho_{\mathcal{A}}\otimes\mathbbm{1}_{\mathcal{B}})) \leqslant \ln d_{\mathcal{A}}$ and $S(\Phi_B(\mathbbm{1}_{\mathcal{A}}\otimes\rho_{\mathcal{B}})) \leqslant \ln d_{\mathcal{B}}$, it is natural to assume that the upper bound for realism-based nonlocality is given by $\eta_{A,B}(\rho) \leqslant \ln(\min\{d_{\mathcal{A}},d_{\mathcal{B}}\})$. This bound is saturated by maximally entangled states, for Schmidt operators $A$ and $B$~\cite{gomes2018}. 

The resource-destroying map assumes here the form $\Lambda_\epsilon^\mathcal{N}(\rho)=(1-\epsilon)\,\rho + \epsilon\,\Phi_{\bar{R}}(\rho)$, with $\bar{R}\in\{\bar{A},\bar{B}\}$. While it is easy to check that $\mathcal{N}(\Lambda_1^\mathcal{N}(\rho))=0$, the proof that $\mathcal{N}(\Lambda_\epsilon^\mathcal{N}(\rho)) \leqslant \mathcal{N}(\rho)$ is more involving~\cite{gomes2019}. Unlike quantum irreality, which, as previously mentioned, turns out to be a useful resource for the distillation of quantum coherence in distributed scenarios~\cite{chitambar2016,streltsov2017}, it is still not clear which quantum processing tasks, if any, realism-based nonlocality can be useful for. In any case, the structure presented here indicates that it suitably fits in our unifying framework of quantum resources. 

\subsection{Further nonclassicalities}

In Ref.~\cite{modi2010}, a perspective is put forward which allows one to quantify a generic correlation, $\mathfrak{Q}$, be it classical or quantum, in terms of the measure $\mathfrak{Q}(\rho)=\min_{\chi\in\mathcal{C}_\mathfrak{Q}}S(\rho||\chi)$, where $\mathcal{C}_\mathfrak{Q}$ is the set of all states $\chi$ such that $\mathfrak{Q}(\chi)=0$. Referring back to the terminology employed in Sect.~\ref{RT}, it is clear that $\chi\equiv\tilde{\rho}$ (free states) and $\mathcal{C}_\mathfrak{Q}\equiv\mathcal{F}$ (free-state set). Now, if the free states admit the form $\tilde{\rho}=\Phi_A(\sigma)$, for a generic state $\sigma$, then we have
\be 
\mathfrak{Q}(\rho)=\min_{\tilde{\rho}\in\mathcal{F}}S(\rho||\tilde{\rho})=\min_{A,\sigma} S(\rho||\Phi_A(\sigma)).
\ee 
The projective nature of the map $\Phi_A$ allows us to show that $\min_\sigma S(\rho||\Phi_A(\sigma))=S(\rho||\Phi_A(\rho))=S(\Phi_A(\rho))-S(\rho)$. It follows that
\be 
\mathfrak{Q}(\rho)=I(\rho)-I(\Phi_{\bar{A}}(\rho)),
\ee 
with $\bar{A}$ being the optimal observable. This means that any nonclassical aspect $\mathfrak{Q}$ admitting $\Phi_A$ as its destroying map will be a resource, at least according to the present approach. In fact, as discussed previously, quantities such as $\mathfrak{Q}(\rho)=\mathcal{R}(\rho)-\mathcal{R}(\Phi_A(\rho))$, for arbitrary resource measures $\mathcal{R}$, will also be resources. Further measures that can be accommodated in our framework are, for example, the measurement-induced disturbance~\cite{luo2008}, 
the quantum dissonance~\cite{modi2010}, and the quintessential coherence~\cite{lami2019,lami2019a}.

\section{Concluding remarks}
This work is intended to contribute to the development of the paradigmatic scenario of resource theories. As a first step, we introduce a class of resource-destroying monitorings, which corresponds to a generic formulation of free operations. These objects form a generalization of resource-destroying maps~\cite{liu2017}, in the sense that they are able to interpolate between resource nongenerating maps and maximally destroying ones. Our second contribution consists of showing that a unified framework can be established that gathers many well-known quantum resources and put them in direct link with information, which then emerges as the most fundamental quantum resource. Specifically, we show that all the studied resources can be viewed as the amount of information that is extracted by an external agent aiming at monitoring some observable(s) of the system. Besides suitably accommodating quantum coherence, entanglement, one-way discord, and symmetric discord in such a unified picture, our approach reveals that the recently proposed quantum irreality and realism-based nonlocality have all the formal characteristics needed for a quantum resource. Our results open up to a larger discussion about the differences among different resources. In our perspective, these differences manifest themselves in terms of the observables about which the external agent decides to get information and establish reality. For future works, it would be interesting to investigate if and how our approach can frame other fundamental resources, such as Bell nonlocality and quantum steering.

\begin{acknowledgements}
A.C.S.C. acknowledges CAPES/Brazil and CNPq/Brazil (Grant Number: 153436/2018-2). R.M.A. acknowledges support from CNPq/Brazil (Grant Number: 303111/2017-8) and the National Institute for Science and Technology of Quantum Information (CNPq, INCT-IQ 465469/2014-0).
\end{acknowledgements}


\begin{thebibliography}{99}

\bibitem{chitambar2019}
E. Chitambar and G. Gour,
Quantum resource theories,
Rev. Mod. Phys. {\bf 91}, 025001 (2019).

\bibitem{horodecki2009}
R. Horodecki, P. Horodecki, M. Horodecki, and K. Horodecki,
Quantum entanglement,
Rev. Mod. Phys. {\bf 81}, 865 (2009).

\bibitem{horodecki2003} 
M. Horodecki, P. Horodecki, and J. Oppenheim, 
Reversible transformations from pure to mixed states and the unique measure of information, 
Phys. Rev. A {\bf 67}, 062104 (2003).

\bibitem{baumgratz2014}
T. Baumgratz, M. Cramer, and M. B. Plenio,
Quantifying coherence,
Phys. Rev. Lett. {\bf 113}, 140401 (2014).

\bibitem{winter2016}
A. Winter and D. Yang,
Operational resource theory of coherence,
Phys. Rev. Lett. {\bf 116}, 120404 (2016).

\bibitem{gour2015}
G. Gour, M. P. M\"uller, V. Narasimhachar, R. W. Spekkens, and N. Y.Halpern,
The resource theory of informational nonequilibrium in thermodynamics,
Phys. Rep. {\bf 583}, 1 (2015).

\bibitem{goold2016}
J. Goold, M. Huber, A. Riera, L. del Rio, and P. Skrzypczyk, 
The role of quantum information in thermodynamics: a topical review,
J. Phys. A {\bf 49}, 143001 (2016).

\bibitem{brandao2015}
F. G. S. L. Brand\~ao and G. Gour,
Reversible framework for quantum resource theories,
Phys. Rev. Lett. {\bf 115}, 070503 (2015).

\bibitem{horodecki2013}
M. Horodecki and J. Oppenheim,
Quantumness in the context of resource theories, 
Int. J. Mod. Phys. B {\bf 27}, 1345019 (2013).

\bibitem{coecke2016}
B. Coecke, T. Fritz, and R. W. Spekkens,
A mathematical theory of resources,
Inf. Comput. {\bf 250}, 59 (2016).

\bibitem{liu2017} 
Z.-W. Liu, X. Hu, and S. Lloyd, 
Resource-destroying maps, 
Phys. Rev. Lett. {\bf 118}, 060502 (2017).

\bibitem{takagi2019}
R. Takagi, B. Regula, K. Bu, Z.-W. Liu, and G. Adesso,
Operational advantage of quantum resources in subchannel discrimination,
Phys. Rev. Lett. {\bf 122}, 140402 (2019).

\bibitem{regula2019}
R. Takagi and B. Regula,
General resource theories in quantum mechanics and beyond: operational characterization via discrimination tasks,
Phys. Rev. X {\bf 9}, 031053 (2019).

\bibitem{bill2015}
A. L. O. Bilobran and R. M. Angelo, 
A measure of physical reality,
Europhys. Letters {\bf 112}, 40005 (2015).

\bibitem{gomes2018}
V. S. Gomes and R. M. Angelo, 
Nonanomalous realism-based measure of nonlocality,
Phys. Rev. A {\bf 97}, 012123 (2018).

\bibitem{gomes2019}
V. S. Gomes and R. M. Angelo,
Resilience of realism-based nonlocality to local disturbance,
Phys. Rev. A {\bf 99}, 012109 (2019).

\bibitem{horodecki2005} 
M. Horodecki, P. Horodecki, R. Horodecki, J. Oppenheim, A. Sen(De), U. Sen, and B. Synak-Radtke, 
Local versus nonlocal information in quantum-information theory: formalism and phenomena, 
Phys. Rev. A {\bf 71}, 062307 (2005).

\bibitem{streltsov2018} 
A. Streltsov, H. Kampermann, S. W\"olk, M. Gessner, and D. Bru\ss, 
Maximal coherence and the resource theory of purity, 
New J. Phys. {\bf 20}, 053058 (2018).

\bibitem{nielsen2000}
M. A. Nielsen and I. L. Chuang,
{\sl Quantum Computation and Quantum Information} 
(Cambridge University Press, Cambridge, 2000).

\bibitem{streltsov2017a}
A. Streltsov, G. Adesso, and M. B. Plenio,
{\sl Colloquium}: quantum coherence as a resource,
Rev. Mod. Phys. {\bf 89}, 041003 (2017).

\bibitem{gisin2002}
N. Gisin, G. Ribordy, W. Tittel, and H. Zbinden,
Quantum cryptography,
Rev. Mod. Phys. {\bf 74}, 145 (2002).

\bibitem{pironio2010}
S. Pironio, A. Ac\'in, S. Massar, A. Boyer de la Giroday, D. N. Matsukevich, P. Maunz, S. Olmschenk, D. Hayes, L. Luo, T. A. Manning, and C. Monroe,
Random numbers certified by Bell's theorem,
Nature {\bf 464}, 1021 (2010).

\bibitem{toth2014}
G. Toth and I. Apellaniz,
Quantum metrology from a quantum information science perspective,
J. Phys. A: Math. Theor. {\bf 47}, 424006 (2014).

\bibitem{bennett1996}
C. H. Bennett, D. P. DiVincenzo, J. A. Smolin, and W. K. Wootters,
Mixed-state entanglement and quantum error correction,
Phys. Rev. A {\bf 54}, 3824 (1996).

\bibitem{wootters1998} 
W. K. Wootters, 
Entanglement of formation of an arbitrary state of two qubits, 
Phys. Rev. Lett. {\bf 80}, 2245 (1998).

\bibitem{zurek2001}
H. Ollivier and W. H. Zurek,
Quantum discord: a measure of the quantumness of correlations,
Phys. Rev. Lett. {\bf 88}, 017901 (2001).

\bibitem{vedral2001}
L. Henderson and V. Vedral,
Classical, quantum and total correlations,
J. Phys. A Math. Gen. {\bf 34}, 6899 (2001).

\bibitem{datta2008}
A. Datta, A. Shaji, and C. M. Caves,
Quantum discord and the power of one qubit,
Phys. Rev. Lett. {\bf 100}, 050502 (2008).

\bibitem{dakic2012}
B. Daki\'c, Y. O. Lipp, X. Ma, M. Ringbauer, S. Kropatschek, S. Barz, T. Paterek, V. Vedral, A. Zeilinger, \v C. Brukner, and P. Walther, 
Quantum discord as resource for remote state preparation,
Nat. Phys. {\bf 8}, 666 (2012).

\bibitem{madhok2013}
V. Madhok and A. Datta, 
Quantum discord as a resource in quantum communication,
Int. J. Mod. Phys. B {\bf 27}, 1345041 (2013). 

\bibitem{pirandola2014}
S. Pirandola, 
Quantum discord as a resource for quantum cryptography,
Sci. Rep. {\bf 4}, 6956 (2014).

\bibitem{rulli2011} 
C. C. Rulli and M. S. Sarandy, 
Global quantum discord in multipartite systems, 
Phys. Rev. A {\bf 84}, 042109 (2011).

\bibitem{xi2012}
Z. Xi, X.-M. Lu, X. Wang, and Y. Li,
Necessary and sufficient condition for saturating the upper bound of quantum discord,
Phys. Rev. A {\bf 85}, 032109 (2012).

\bibitem{xi2011}
Z. Xi, X.-M. Lu, X. Wang, and Y. Li,
The upper bound and continuity of quantum discord,
J. Phys. A: Math. Theor. {\bf 44}, 375301 (2011).

\bibitem{dieguez2018} 
P. R. Dieguez and R. M. Angelo, 
Weak quantum discord, 
Quantum Inf. Process. {\bf 17}, 194 (2018).

\bibitem{dieguez2018a}
P. R. Dieguez and R. M. Angelo,
Information-reality complementarity: the role of measurements and quantum reference frames,
Phys. Rev. A {\bf 97}, 022107 (2018).

\bibitem{freire2019}
I. S. Freire and R. M. Angelo, 
Quantifying continuous-variable realism,
Phys. Rev. A {\bf 100}, 022105 (2019).

\bibitem{orthey2019}
A. C. Orthey and R. M. Angelo,
Nonlocality, quantum correlations, and violations of classical realism in the dynamics of two noninteracting quantum walkers,
Phys. Rev. A {\bf 100}, 042110 (2019).

\bibitem{chitambar2016}
E. Chitambar, A. Streltsov, S. Rana, M. N. Bera, G. Adesso, and M. Lewenstein,
Assisted distillation of quantum coherence,
Phys. Rev. Lett. {\bf 116}, 070402 (2016).

\bibitem{streltsov2017}
A. Streltsov, S. Rana, M. N. Bera, and M. Lewenstein,
Towards resource theory of coherence in distributed scenarios,
Phys. Rev. X {\bf 7}, 011024 (2017).

\bibitem{zurek2003}
W. H. Zurek,
Quantum discord and Maxwell's demons,
Phys. Rev. A {\bf 67}, 012320 (2003).

\bibitem{costa2013}
A. C. S. Costa and R. M. Angelo,
Bayes' rule, generalized discord, and nonextensive thermodynamics,
Phys. Rev. A {\bf 87}, 032109 (2013).

\bibitem{modi2010}
K. Modi, T. Paterek, W. Son, V. Vedral, and M. Williamson,
Unified view of quantum and classical correlations,
Phys. Rev. Lett. {\bf 104}, 080501 (2010).

\bibitem{luo2008}
S. Luo,
Using measurement-induced disturbance to characterize correlations as classical or quantum,
Phys. Rev. A {\bf 77}, 022301 (2008).

\bibitem{lami2019}
L. Lami, B. Regula, and G. Adesso,
Generic bound coherence under strictly incoherent operations,
Phys. Rev. Lett. {\bf 122}, 150402 (2019).

\bibitem{lami2019a}
L. Lami,
Completing the grand tour of asymptotic quantum coherence manipulation,
IEEE Trans. Inf. Theory {\bf 66}, 2165 (2019).




\end{thebibliography}
\end{document}